\documentclass[11pt]{article}
\usepackage{moriond}

\usepackage{amsfonts}
\usepackage[english]{babel}
\usepackage{graphicx,url}
\usepackage[colorlinks=true,urlcolor=blue,linkcolor=black]{hyperref}
\def\invfb{fb^{-1}}

\newcommand{\Zgamgam}{{\color{black}4.0}}
\newcommand{\Zfourlepton}{{\color{black}4.4}}
\newcommand{\Zww}{{\color{black}3.0}}
\newcommand{\Zbb}{{\color{black}1.8}}
\newcommand{\Ztautau}{{\color{black}1.8}}
\newcommand{\ZhighRes}{{\color{black}5.8}}

\newcommand{\MaxLocalZ}{{\color{black}6.9}}

\newcommand{\expZgamgam}{{\color{black}2.8}}
\newcommand{\expZfourlepton}{{\color{black}5.0}}
\newcommand{\expZww}{{\color{black}4.3}}
\newcommand{\expZbb}{{\color{black}2.2}}
\newcommand{\expZtautau}{{\color{black}2.1}}
\newcommand{\expZhighRes}{{\color{black}5.7}}

\newcommand{\expMaxLocalZ}{{\color{black}7.8}}

\newcommand{\MaxZmass}{{\color{black}125.8}}    

\newcommand{\MUHAT}{{\color{black}$0.88 \pm 0.21$}}  

\newcommand{\MASSstat}{{\color{black}0.4}} 
\newcommand{\MASSsyst}{{\color{black}0.4}} 
\newcommand{\MASS}{{\color{black}$125.8~\pm \MASSstat~(\mathrm{stat})~\pm \MASSsyst~(\mathrm{syst})$}} 
\newcommand{\MASSH}{{\color{black}$125.8~ \pm 0.5$}}    
\newcommand{\mX}{{\color{black}125.8}}                  

\newcommand{\lwzONE}{{\color{black}[0.57--1.65]}}        
\newcommand{\BRBSM}{{\color{black}[0.00--0.62]}}           
\newcommand{\llq}{{\color{black}[0.00--2.11]}}             
\newcommand{\ldu}{{\color{black}[0.45--1.66]}}             

\newcommand{\LumiEight}{{\color{black}12.2}}   

\newcommand{\PH}{\mathrm{H}}
\newcommand{\mH}{\ensuremath{m_{\mathrm{H}}}}
\newcommand{\CLs}{\ensuremath{\mathrm{CL_s}}}

\newcommand{\Pgg}{\gamma}
\newcommand{\Pgt}{\tau}
\newcommand{\cPqb}{\mathrm{b}}
\newcommand{\PW}{\mathrm{W}}
\newcommand{\cPZ}{\mathrm{Z}}
\newcommand{\cPgn}{\nu}

\begin{document}
\vspace*{4cm}
\title{Combination and Standar Model Scalar Boson Properties in CMS}

\author{Mingshui Chen for the CMS Collaboration}

\address{University of Florida, Gainesville, USA}

\maketitle\abstracts{Combination results of the recently discovered boson are presented using data samples
corresponding to integrated luminosities of up to 5.1 $\invfb$ at 7 TeV and 
up to 12.2 $\invfb$ at 8 TeV of proton-proton collisions collected with CMS experiment at LHC.   
The significance of the new boson is 6.9 $\sigma$ with mass measured to be 125.8 $\pm$ 0.4 (stat) $\pm$ 0.4 (syst). 
The event yields obtained by the different analyses targeting specific decay modes and production mechanisms are consistent with those 
predicted for the stand model (SM) Higgs boson. The best-fit signal strength for all channels combined,
 expressed in units of the SM Higgs boson cross section, is 0.88 $\pm$ 0.21 at the measured mass. 
The consistency of the couplins of the observed boson with those expected for the SM Higgs boson is tested in various ways, 
and no significant deviations are found. 
Results on the test of different spin-parity hypotheses of the observed boson are also shown, 
but with updated data samples corresponding to integrated luminosities of 5.1 $\invfb$ at 7 TeV and 
19.6 $\invfb$ at 8 TeV in two channels H $\rightarrow$ WW $\rightarrow~2\ell2\nu$ and H $\rightarrow$ ZZ $\rightarrow~4\ell$ separately.  
Under the assumption that the observed boson has spin 0 and positive parity, 
the pure scalar hypothesis is found to be consistent with the observed boson when compared to other tested spin-parity hypotheses.
 The data in the H $\rightarrow$ ZZ $\rightarrow~4\ell$ channel disfavor the pseudo-scalar hypothesis $0^-$ with a CLs value of 0.16$\%$, disfavor the pure spin-2 hypothesis of a narrow
resonance with the minimal couplings to the vector bosons with a CLs value of 1.5$\%$, and disfavor the pure spin-1 hypothesis with even smaller CLs value.}

\section{Introduction}
One of the primary goals at the Large Hadron Collider (LHC) is to understand the mechanism for electroweak symmetry breaking.
To achieve the symmetry breaking, a complex scalar doublet is introduced in the standard model~\cite{Glashow:1961tr,Weinberg:1967tq,sm_salam}, 
leading to  the prediction of the Higgs
boson ($\PH$)~\cite{Englert:1964et,Higgs:1964ia,Higgs:1964pj,Guralnik:1964eu,Higgs:1966ev,Kibble:1967sv}.

In this proceeding we report on the results from the searches for the SM Higgs boson 
and the measurements of the properties of the recently observed boson with a mass near 125 GeV by 
CMS~\cite{CMSobservation125} and ATLAS~\cite{ATLASobservation125}. These measurements are carried out in proton-proton collisions at $\sqrt{s}=7$ (2011 data) and 8 TeV
(2012 data) using the Compact Muon Solenoid (CMS) detector~\cite{CMS:2008zzk} at the LHC. 
All results presented here are obtained with data analysed corresponding to integrated luminosities of up to 5.1~$\invfb$ at 7 TeV and $\LumiEight ~\invfb$ at 8 TeV~\cite{HIG-12-045}, 
 except for the test of spin-parity hypotheses of the observed boson which uses the full data corresponding to 
integrated luminosities of up to 5.1~$\invfb$ at 7 TeV and $19.6 ~\invfb$ at 8 TeV~\cite{HIG-13-002,HIG-13-003}.

The CMS detector~\cite{bib-detector} consists of a barrel assembly and two endcaps,
comprising, in successive layers outwards from the collision region,
the silicon pixel and strip tracker,
the lead tungstate crystal electromagnetic calorimeter,
the brass/scintillator hadron calorimeter,
the superconducting solenoid,
and gas-ionization chambers embedded in the steel return yoke for the detection of muons.

This proceeding is organized as follows. 
Section 2 briefly describes the production and decay modes relevant for the search channels went into this combination. 
Section 3 gives the concise  definitions of statistical quantities 
we use for characterizing the outcome of the search. Results and summaries are presented in section 4 and 5 respectively.    

\section{Search channels}

In pp collisions at $\sqrt{s}=$ 7-8 TeV, the SM Higgs boson production is dominated by the gluon-gluon fusion mode. 
There are also other relevant production modes :  vector boson fusion (VBF),
associated $\mathrm{WH}$ and $\mathrm{ZH}$ production,
and production in association with top quarks, $t\bar{t}\mathrm{H}$.
The relevant decay modes of the SM Higgs boson around 125 GeV are the following:
$\PH \to \Pgg\Pgg$,
$\PH \to \Pgt^+\Pgt^-$ (denoted as $\PH \to \Pgt\Pgt$),
followed by leptonic and hadronic decays of the $\Pgt$-leptons,
  $\PH \to b\bar{b}$ (denoted as $\PH \to \cPqb\cPqb$),
$\PH \to \PW\PW$,
followed by $\PW\PW \to \ell\cPgn\ell\cPgn$, 
and $\PH\to\cPZ\cPZ$,
followed by $\cPZ\cPZ$ decays to
$4\ell$.
Here and throughout, $\ell$ stands for electrons or muons.
The cross section and decay branching fractions of the SM Higgs boson, together with their uncertainties, 
are taken from Ref.~\cite{LHCHiggsCrossSectionWorkingGroup:2011ti,Dittmaier:2012vm,LHC-HCSG}.
The total cross section at $\sqrt{s}=$7 (8) TeV varies from 23 (29) to 15 (19) pb for a Higgs boson mass range from 110 to 135 GeV. 

Around 125 GeV, the $\PH \to \Pgg\Pgg$ and $\PH\to\cPZ\cPZ \to 4\ell$
channels have the best sensitivity due to the excellent mass resolution for the reconstructed diphoton and four-lepton
final states, respectively. The $\PH\to\cPZ\cPZ \to 4\ell$
channel also benefits from low backgroud level.
The $\PH \to \PW\PW \to \ell\cPgn\ell\cPgn$ channel also has high sensitivity,
but has relatively poor mass resolution due to the neutrinos in the final state.
The $\cPqb\cPqb$ and $\Pgt\Pgt$ channels have large backgrounds and poor mass resolutions, hence less sensitivity.

The results presented in this proceeding are obtained by combining Higgs boson searches
exploiting different production and decay modes. Table~\ref{tab:ProductionDecay}
shows modes used in the searches. More detailed descriptions of all search analyses used in the combination
can be found in Refs.~\cite{HIG-12-015,HIG-12-041,HIG-12-042,HIG-12-043,HIG-12-044} 

\begin{table} [h]
\begin{center}
\small
\caption{Summary of production mechanisms and decay channels explicitly targeted
in searches for a low mass Higgs boson ($m_{\PH}<135$ GeV).
``Inclusive'' searches include gluon-gluon fusion $gg \to \PH$
plus any phase space not covered by searches targeting VBF, V$\PH$ (V stands for $\PW$ or $\cPZ$),
and  $t\bar{t}\PH$ production. All analyses targeting particular production
mechanism have admixture, sometimes very substantial, of other mechnisms. }
\label{tab:ProductionDecay}
\begin{tabular}{|l||c|c|c|c|}
\hline
                          & ``inclusive''  & VBF          & V$\PH$          & ttH  \\
\hline\hline
$ \PH \to \gamma\gamma$   &   \checkmark   &  \checkmark  &               &      \\
\hline
$ \PH \to bb$             &                &              &  \checkmark   &  \checkmark   \\
\hline
$ \PH \to \tau\tau$       &   \checkmark   &  \checkmark  &  \checkmark   &      \\
\hline
$ \PH \to WW$             &   \checkmark   &  \checkmark  &  \checkmark   &      \\
\hline
$ \PH \to \cPZ\cPZ$       &   \checkmark   &              &               &      \\
\hline
\end{tabular}
\end{center}
\end{table}

\section{Combination methodology}
The description of the overall statistical methodology can be found in Refs.~\cite{LHC-HCG-Report, Chatrchyan:2012tx}. 
Below we give brief definitions of statistical quantities 
we use for characterizing the outcome of the search. 
Results presented in this proceeding are obtained using asymptotic formulae~\cite{Cowan:2010st}.

\subsection{Characterising an excess of events: {\it p}-values and significance}

To quantify the inconsistency of the observed excess with the background-only hypothesis, 
we use the statistical significance $Z$ for a signal-like excess which is computed from the probability $p_0$ 
(known as the $p$-value, using the one-sided Gaussian tail convention): 

\begin{equation}
p_0 = \mathrm{P}(q_0 \geq q_0^{obs} \, | \, \mathrm{b}),
\end{equation} 

\begin{equation}
\label{eq:Z}
p_0 \, = \, \int_{Z}^{+\infty} \frac{1}{\sqrt{2\pi}} \exp(-x^2/2) \,\, \mathrm{d}x.
\end{equation}

where $q_0$ is a test statistic based on the profile likelihood ratio defined as below:

\begin{equation}
\label{eq:method_q0}
 q_{0} \, = \, - 2 \, \ln \frac {\mathcal{L}(\mathrm{obs} \, | \, b, \, \hat \theta_{0} ) }
                       {\mathcal{L}(\mathrm{obs} \, | \, \hat \mu \cdot s + b, \, \hat \theta ) } ,
\end{equation}

where $s$ stands for the signal expected under the SM Higgs hypothesis, 
$\mu$ is a signal strength modifier
introduced to accommodate deviations from SM Higgs predictions,
$b$ stands for backgrounds, and $\theta$ are nuisance parameters 
describing systematic uncertainties 
($\hat \theta_{0}$ maximizes the likelihood in the numerator for background-only hypothesis,
while $\hat \mu$ and $\hat \theta$ define the point at which the likelihood reaches its global maximum).

Systematic uncertainties are incorporated in the analysis via nuisance parameters and
are treated according to the frequentist paradigm.

\subsection{Extracting signal model parameters}

Signal model parameters $a$ (signal strength modifier $\mu$ can be one of them)
are evaluated from a scan of the profile likelihood ratio $q(a)$:

\begin{equation}
q(a) = \, - 2 \, \ln \frac {\mathcal{L}(\mathrm{obs} \, | \, s(a) + b,      \, \hat \theta_{a} ) }
                      {\mathcal{L}(\mathrm{obs} \, | \, s(\hat a) + b, \, \hat \theta ) } ,
\end{equation}

Parameters $\hat a$ and $\hat \theta$ that maximize the likelihood, 
$\mathcal{L}(\mathrm{obs} \, | \, s(\hat a) + b, \, \hat \theta ) = \mathcal{L}_{\mathrm{max}}$, 
are called the best-fit set. 
The 68\%~(95\%)~CL on a given parameter of interest $a_i$ is evaluated from $q(a_i)=1$~(3.84) 
with all other unconstrained model parameters treated in the same way as the nuisance parameters. 
The 2D 68\%~(95\%)~CL contours for pairs of parameters are derived from $q(a_i, a_j) = 2.3$~(6).

\section{Results}

\subsection{Significance of the observed excess}
\label{sec:significance}

\begin{figure*} 
\centering
\includegraphics[width=0.44\textwidth]{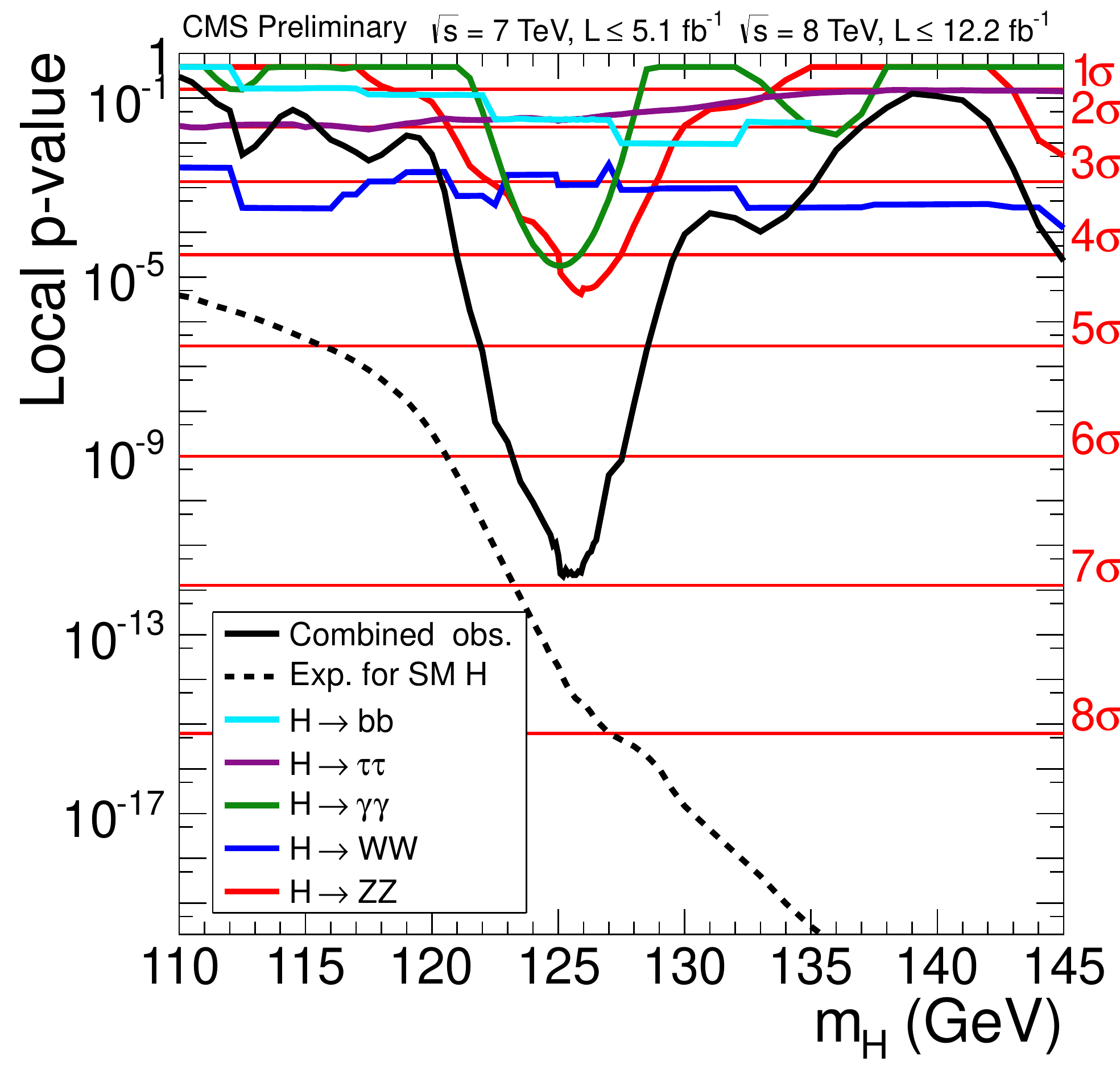}  \hfill
\includegraphics[width=0.44\textwidth]{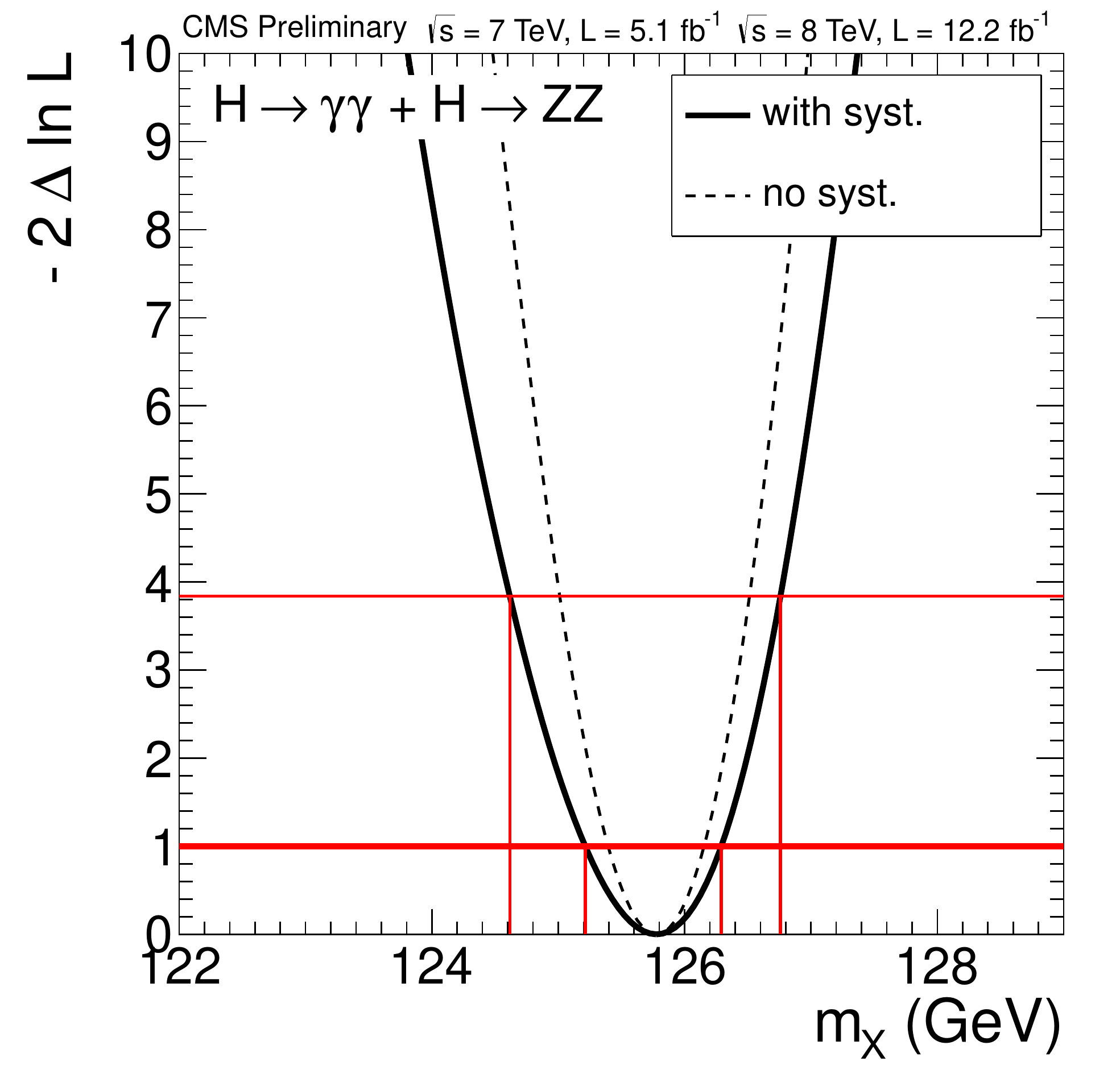} 
\caption{
(Left) The observed local $p$-value $p_0$ for five sub-combinations by decay mode
and the overall combination as a function of the Higgs boson mass
in the range 110--145~GeV. 
The dashed lines show the expected local $p$-values $p_0(\mH)$,
should a SM Higgs boson with mass $\mH$ exist.
(Right) 1D-scans of the test statistic $q(m_{\mathrm{X}})$ versus hypothesised boson mass $m_{\mathrm{X}}$
for the combination of the $\gamma \gamma$ and 4$\ell$ final states.
The solid line is obtained with all nuisance parameters profiled
and, hence, includes both statistical and systematic uncertainties.
The dashed line is obtained with all nuisance parameters fixed to their best-fit values
and, hence, includes only statistical uncertainties.
The crossings with the thick (thin) horizontal lines define the 68\% (95\%) CL interval
for the measured mass.
}
\label{fig:pvalue_by_decay}
\end{figure*}

Fig.~\ref{fig:pvalue_by_decay} (left) shows the local $p$-values for the various sub-combinations by decay channel and for the overall combination.
The largest significance is $\MaxLocalZ \, \sigma$ at the mass \MaxZmass~GeV with the dataset used in this combination,
which confirms the previous observation of the new boson.
 The largest contributors to the overall excess in the combination
near the mass of 125~GeV are the $\cPZ\cPZ \to 4\ell$ and  $\Pgg\Pgg$ channels,
with maximum significances of $\Zfourlepton \, \sigma$ and $\Zgamgam \, \sigma$.
The $\PW\PW$ channel contributes to about $3\, \sigma$, and the bb and $\tau\tau$ contribute about $2\, \sigma$ each.

Table~\ref{tab:Signif} summarises the median expected and observed local significances
for a SM Higgs boson mass hypothesis of \MaxZmass~GeV
for the individual decay modes and their various combinations.
The expected significance is evaluated for
a pseudo-observation equal to the expected background and signal rate.
The $\pm 1 \sigma$ range around the most probable significance
should contain 68\% of the statistical fluctuations that could occur in data.

\begin{table}[htbp]
\begin{center}
\caption{
The significance of the median expected and observed event excesses
in individual decay modes and their various combinations
for a SM Higgs boson mass hypothesis of \MaxZmass~GeV.
}
\label{tab:Signif}
\begin{tabular}{l|c|c}
\hline
Decay mode or combination & Expected ($\sigma$) & Observed ($\sigma$) \\
\hline
$\cPZ\cPZ$    &  \expZfourlepton  & \Zfourlepton  \\ %
$\Pgg\Pgg$    &  \expZgamgam       & \Zgamgam  \\ %
$\PW\PW$      &  \expZww           & \Zww  \\ %
$\cPqb\cPqb$  &  \expZbb           & \Zbb  \\ %
$\Pgt\Pgt$    &  \expZtautau       & \Ztautau  \\ %
\hline
$\Pgg\Pgg$ + $\cPZ\cPZ$                 & \expZhighRes  & \ZhighRes \\
\hline
$\Pgg\Pgg$ + $\cPZ\cPZ$ + $\PW\PW$ + $\Pgt\Pgt$ + $\cPqb\cPqb$ & \expMaxLocalZ & \MaxLocalZ \\%
\hline
\end{tabular}
\end{center}
\end{table}

\subsection{Mass of the observed state}
\label{sec:mass}

In this measurement, 
we use the $\cPZ\cPZ \to 4\ell$ and $\gamma\gamma$ channels
that have excellent mass resolution ($1-2\%$).
The signal in all channels is assumed to arise from a state with mass $m_{\mathrm{X}}$.   
We extract the mass $m_{\mathrm{X}}$ and its uncertainty from a scan of the combined
test statistic $q(m_{\mathrm{X}})$  with independent 
signal strength modifiers for the $ gg \to \PH \to \gamma\gamma$,
VBF+VH$\to \gamma\gamma$, and $\PH \to \cPZ\cPZ \to 4\ell$
processes separately. The three signal signal strength modifiers are
profiled in the same way as all other nuisance parameters.
Figure~\ref{fig:pvalue_by_decay} (right) shows the scan of the test statistic
as a function of the hypothesised mass $m_{\mathrm{X}}$. The solid curve is with all nuisance parameters profiled, while the 
dashed curve is with all nuisance parameters fixed to their best-fit values.
Crossings of the $q(m_{\mathrm{X}})$ curves with horizontal thick (thin) lines at 1 (3.8) define
the 68\% (95\%) CL intervals for the mass of the observed particle.  
The intervals with solid curve include
both statistical and systematic uncertainties.
The 68\% CL interval is \mbox{$m_{\mathrm{X}} = $\MASSH~GeV}~(stat+syst).
The intervals with dashed line define the statistical error (68\% CL interval) in the
mass measurement: $m_{\mathrm{X}} = \mX \pm \MASSstat $~(stat.)~GeV.
The final mass measurement can be written as $m_{\mathrm{X}} = $~\MASS~GeV.

\subsection{Signal strength in combination and sub-combinations}
\begin{figure*} [b]
\centering
\includegraphics[width=0.44\textwidth]{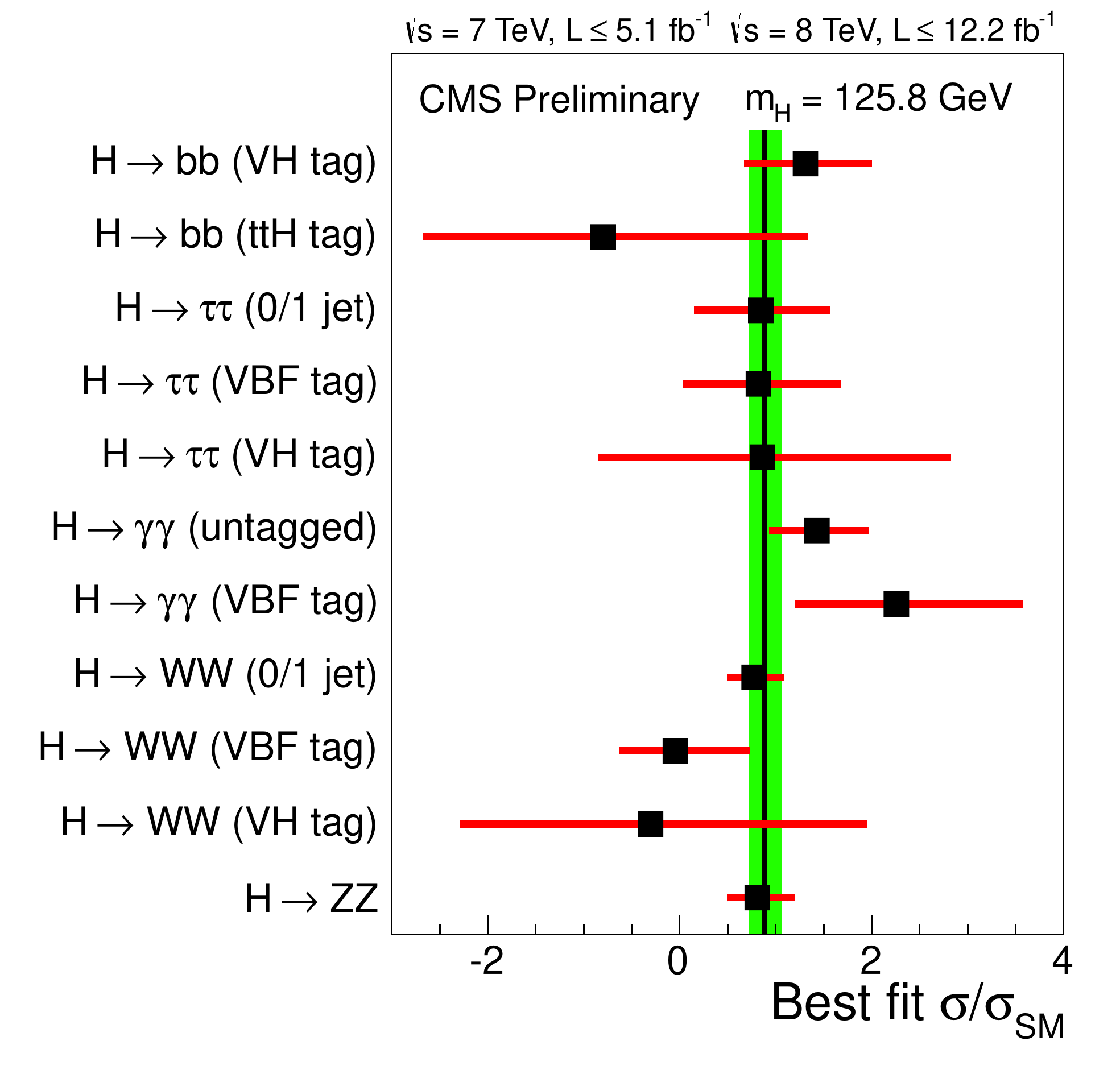} \\
\caption{
 	Best-fit signal strength values $\hat \mu = \sigma / \sigma_\mathrm{SM}$
	for various sub-combinations by decay mode and by additional tags
      targeting a particular production mechanism.
	The horizontal bars indicate the $\pm 1 \sigma$ total (statistical plus systematic) uncertainties.
	The solid vertical line with the band shows the overall combined $\hat \mu$ value with its uncertainties.
    }
\label{fig:muhat_compatibility}
\end{figure*}

Figure~\ref{fig:muhat_compatibility} shows the best fit value for the signal strength modifier 
 $\hat \mu = \sigma / \sigma_{\mathrm{SM}}$ in data,
obtained in the combination of all search channels,
as well as in different sub-combinations of search channels 
for \mH = \mX~GeV, organized by decay mode and by additional tags used to select preferentially events from a particular production mechanism
(Note that the expected purities of the different tagged samples vary substantially).
The observed $\hat \mu$ value of the full combination 
for a hypothesised Higgs boson mass of \mX~GeV is found to be \MUHAT\
and is consistent with the value expected for the SM Higgs boson ($\mu =1$)
within the $\pm 1 \sigma$ uncertainties (statistical+systematic). 
None of the sub-combinations depart from the SM Higgs boson hypothesis, $\mu = 1$,
by a significant deviation with respect to their current individual sensitivities.

\subsection{Compatibility of the observed data with the SM Higgs boson couplings}

In this section, we follow the prescriptions of the LHC Higgs Cross section working group
\cite{LHCHiggsCrossSectionWorkingGroup:2012nn}, implement the couplings compatibility tests with the following assumptions:
\begin{itemize}
\item The signals observed in the different search channels are due to a single narrow resonance.
\item Zero-width approximation of this observed state is used.
\item The observed state is assumed to be a CP-even scalar as in the SM.
\end{itemize} 

For a (production)$\times$(decay) mode, its event yield is proportional to its production cross section and partial decay width. 
We introduce scale factors $\kappa_i$ to scale the cross section $\sigma_i$ and the partial decay widths $\Gamma_i$ associated with the
SM particle $i$ by comparing to the corresponding SM prediction as following equation: 

\begin{equation}
N(xx \to \PH \to yy) \sim \sigma(xx \to \PH) \cdot \mathcal{B}(\PH \to yy)
                     \sim \frac {\Gamma_{xx} \, \Gamma_{yy} }
                                { \Gamma_{\mathrm{tot}} } \sim \frac {\kappa^{2}_{x}\,\kappa^{2}_{y}} {\Gamma_{\mathrm{tot}}}.
\end{equation}

For current searches, there are seven scale factors (corresponding to seven partial widths related to  
 W, Z, top quark, b quark, $\tau$, gluon and $\gamma$) and the total width which are relevant. 
The $gg\to \PH$ and $\PH \to \gamma\gamma$ are loop induced in the SM and are directly sensitive to the presence of new physics.
The possibility of Higgs boson decays to beyond-standard-model (BSM) particles,
with a partial width $\Gamma_{\mathrm{BSM}}$, is
accommodated by keeping $\Gamma_{\mathrm{tot}}$ as an independent parameter so that
$\Gamma_{\mathrm{tot}} = \sum \Gamma_{i (\mathrm{SM})} + \Gamma_{\mathrm{BSM}}$,
where $\Gamma_{ i (\mathrm{SM})}$ stands for the partial widths of decays to SM particles.

With current dataset, we present a number of combinations with a more limited number of degrees of freedom 
instead of extracting all eight parameters.
The remaining un-measured degrees of freedom are either constrained to be equal to the SM Higgs boson expectations
or profiled in the likelihood scans together with all other nuisance parameters.

$\ $

\textit{Test of the custodial symmetry}

The $\mathrm{SU(2)_L}$ custodial symmetry~\cite{Sikivie:1980hm} requires identical coupling scale factors for W and Z bosons, $\kappa_W$ and $\kappa_Z$. 
To test this custodial symmetry, we probe the consistency of the ratio $\lambda_{\mathrm{wz}} = \kappa_{\mathrm{w}} / \kappa_{\mathrm{z}}$ with unity.

For this test, the results presented here use both the inclusive 
$pp \to \PH \to \cPZ\cPZ$
and untagged $pp \to \PH \to \PW\PW$ search channels. 
The dominant production contribution to the two channels is $gg \to \PH$. Therefore,
the ratio of event yields in these channels provides
a nearly model independent measurement of $\lambda_{\mathrm{wz}}$.
The free  parameters in this test are 
 $\kappa_{\mathrm{z}}$ and $\lambda_{\mathrm{wz}}$.
 $\kappa_{\mathrm{z}}$ is treated as
a nuisance parameter, while all fermionic scale factors $\kappa_F$ =1.
A likelihood scan vs $\lambda_{\mathrm{wz}}$ is performed and the 95\%~CL interval for $\lambda_{\mathrm{wz}}$ is \lwzONE. 
The data are consistent with the SM expectation ($\lambda_{\mathrm{wz}}=1$).

In all combinations presented further, we assume $\lambda_{\mathrm{wz}}=1$ and use a common factor $\kappa_{\mathrm{V}}$
to modify the couplings to $\PW$ and $\cPZ$ bosons, whilst preserving their ratio.

$\ $

\textit{Test for asymmetries in couplings to fermions}

In models with two Higgs doublets (2HDM), the couplings of the neutral Higgs bosons to fermions
can be substantially modified with respect to the Yukawa couplings of the SM Higgs boson. In more
general 2HDMs, leptons can be made to virtually decouple from the Higgs boson that otherwise behaves
in a SM like way with respect to $\PW$/$\cPZ$-bosons and quarks.
Inspired by the possibility of such modifications to the fermion couplings,
we perform two combinations, in which we allow for
different ratios of the couplings to down/up fermions
($\lambda_{\mathrm{du}} = \kappa_{\mathrm{d}} / \kappa_{\mathrm{u}}$)
or different ratios of couplings to lepton and quarks
($\lambda_{\ell\mathrm{q}} = \kappa_{\ell} / \kappa_{\mathrm{q}}$).
We assume that $\Gamma_{\mathrm{BSM}}=0$.
Both $\lambda_{\mathrm{du}}$ and $\lambda_{\ell \mathrm{q}}$ are constrained to
be positive; the 95\% CL intervals for them are \ldu\ and \llq\, respectively.

$\ $

\textit{Test of couplings to the vector bosons and fermions}

This test assumes that $\Gamma_{\mathrm{BSM}}=0$, i.e. no new Higgs boson decay modes are open.
The free parameters are the coupling scale factors $\kappa_\mathrm{V}$ for all vector boson and $\kappa_F$ for all fermion couplings.
At LO, all partial widths,
except for $\Gamma_{\Pgg\Pgg}$, scale either as
$\kappa^2_\mathrm{V}$ or $\kappa^2_F$.
 The partial width $\Gamma_{\Pgg\Pgg}$ is induced via
$\PW$ and top loop diagrams and scales as $| \alpha \, \kappa_\mathrm{V} + \beta \, \kappa_F |^2$.

\begin{figure*}[bhtp]
\centering
\resizebox{!}{0.44\textwidth}{\includegraphics{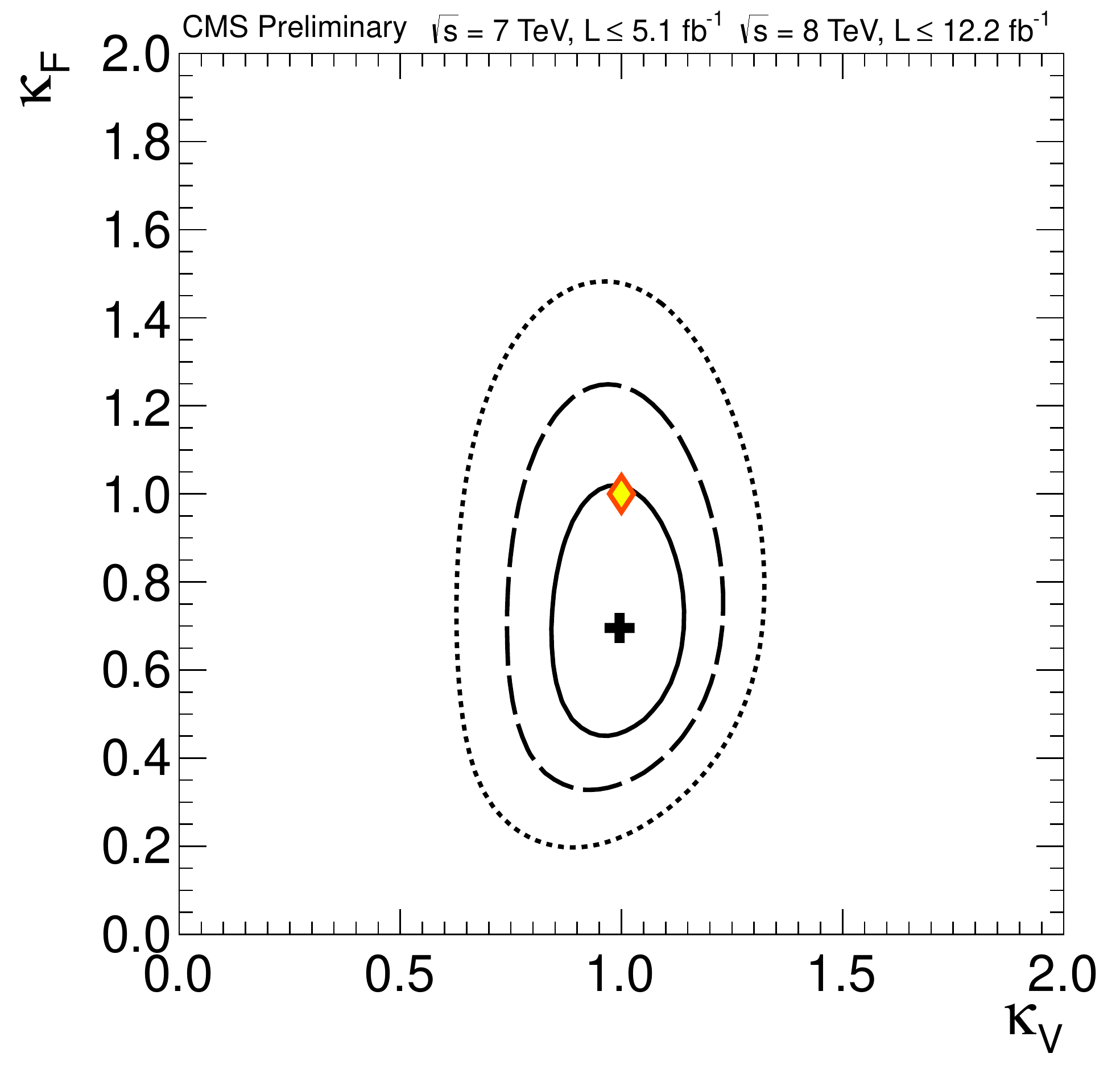}}\hfill
\resizebox{!}{0.44\textwidth}{\includegraphics{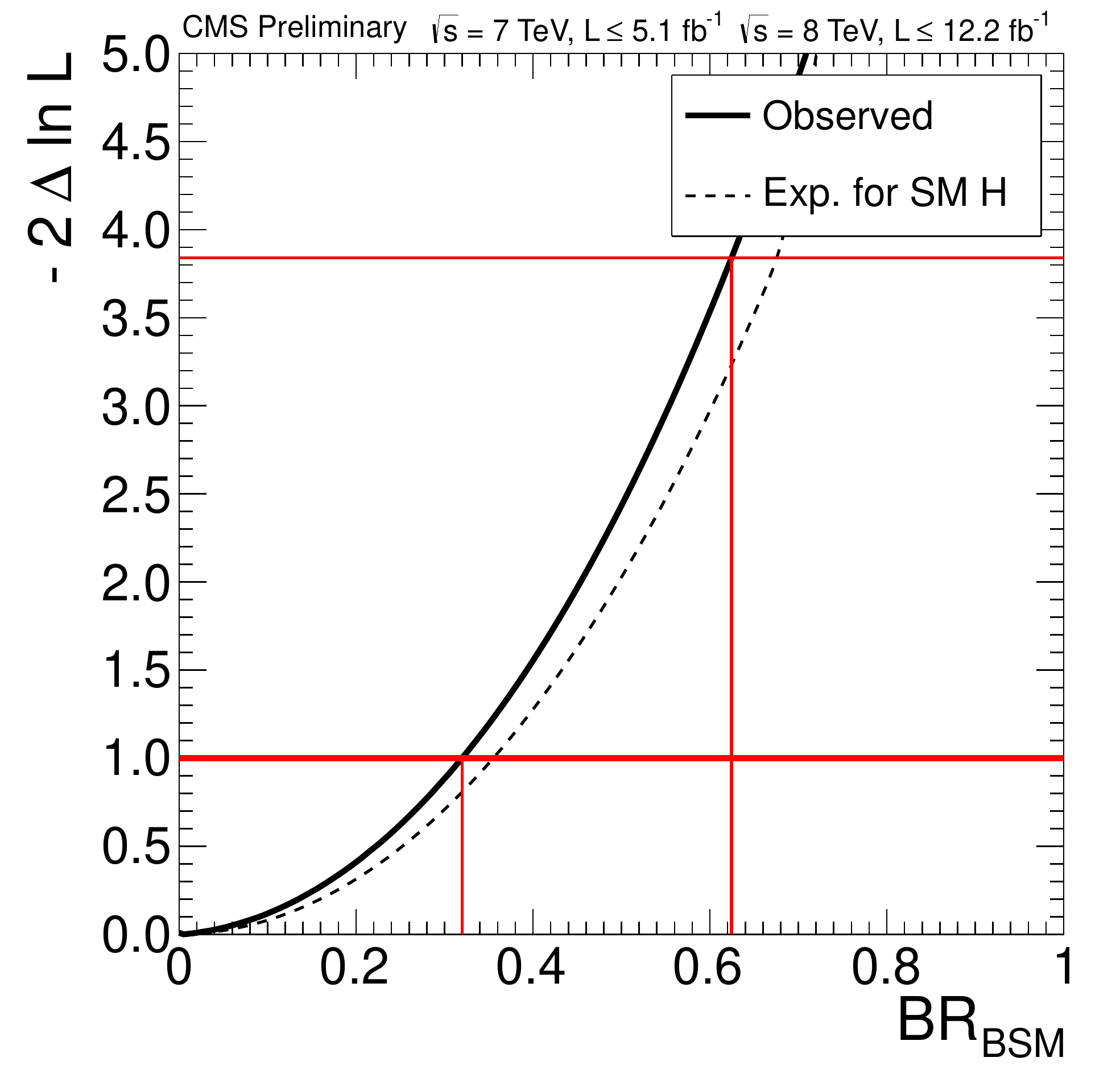}}
\caption{
(Left) The 2D-scan of the likelihood (test statistic) $q(\kappa_{\mathrm{V}},\kappa_F)$
vs the $(\kappa_{\mathrm{F}}, \kappa_F)$ parameters.
The cross indicates the best-fit values. The solid, dashed and dotted contours show the
68\%, 95\% and 99.7\%CL ranges, respectively. The yellow diamond shows the SM point
$(\kappa_{\mathrm{V}}, \kappa_F)=(1,1)$. 
(Right) The likelihood scan versus $\mathrm{BR}_{\mathrm{BSM}}=\Gamma_{\mathrm{BSM}}/\Gamma_{\mathrm{tot}}$,
The solid curve is the data and the dashed line indicates the expected median results in the presence of the SM Higgs boson.
The partial widths associated with the tree-level production processes
and decay modes are assumed to be unaltered ($\kappa = 1$). 
}
\label{fig:BSM2}
\end{figure*}

Figure~\ref{fig:BSM2} shows the 2D likelihood scan
over the $(\kappa_{\mathrm{V}},\,\kappa_F)$ phase space.
The 68\%, 95\% and 99.7\% confidence regions for $\kappa_{\mathrm{V}}$ and $\kappa_F$
are shown with solid, dashed and dotted lines, respectively. The data are compatible with the expectation for the
standard model Higgs boson: the point ($\kappa_V,\kappa_F$)=(1,1) is within the 95\% confidence interval defined by data.

$\ $

\textit{Test for presence of BSM particles}

New particles could contribute in loops or in new final states. 
The free parameters in this test are effective scale factors $\kappa_g$, $\kappa_\gamma$ and
 branching fraction $\mathrm{BR}_{\mathrm{BSM}}=\Gamma_{\mathrm{BSM}}/\Gamma_{\mathrm{tot}}$.

Figure~\ref{fig:BSM2} shows the likelihood scan versus
$\mathrm{BR}_{\mathrm{BSM}}$,
while $\kappa_g$ and $\kappa_{\Pgg}$ are profiled together with all other nuisance parameters.
The partial widths associated with the tree-level production processes
and decay modes are assumed to be unaltered ($\kappa = 1$).
The 95\% CL inerval for $\mathrm{BR}_{\mathrm{BSM}}$  is \BRBSM, which shows no sign of new physics.


More coupling tests can be found in Ref.~\cite{HIG-12-045}.

\subsection{Test of different spin-parity hypotheses}
\label{sec:JCP}

In a recent CMS publication~\cite{Chatrchyan:2012jja}, the data are found to be consitent with the pure scalar hypothesis
and disfavor the pure pseudo-scalar hypothesis with the assumption that the observed state is spin 0.  
With the full 2011+2012 dataset, we perform a more comprehensive set of hypothesis tests between the SM Higgs boson and other  
signal models with $J^P=0_\mathrm{h}^+$, $0^-$, $2_\mathrm{m}^+$(gg), $2_\mathrm{m}^+$($\mathrm{q\bar{q}}$), $1^-$, and $1^+$ in the same channel~\cite{HIG-13-002}.
We also perform the hypothesis test between $0^+$ and $2_\mathrm{m}^+$(gg)
in the $\PH\to \PW\PW \to \ell\nu\ell\nu$ channel~\cite{HIG-13-003}.
The exact definition of the coupling structure of
the alternative states can be found in table~\ref{tab:jpmodels} and in Ref.~\cite{Bolognesi:2012mm}.

These hypothesis tests are based on the following test statistic
\begin{equation}
q=-2\,{\ln({\cal L}_{J^P}/{\cal L}_{SM})} 
\end{equation}
where the likelihood corresponding to the background plus SM $0^+$ (alternative $J^P$ signal model) is denoted by ${\cal L}_{SM}$ (${\cal L}_{J^P}$). 

\begin{figure}[th!]
\begin{center}
\centerline{
\includegraphics[width=0.32\linewidth]{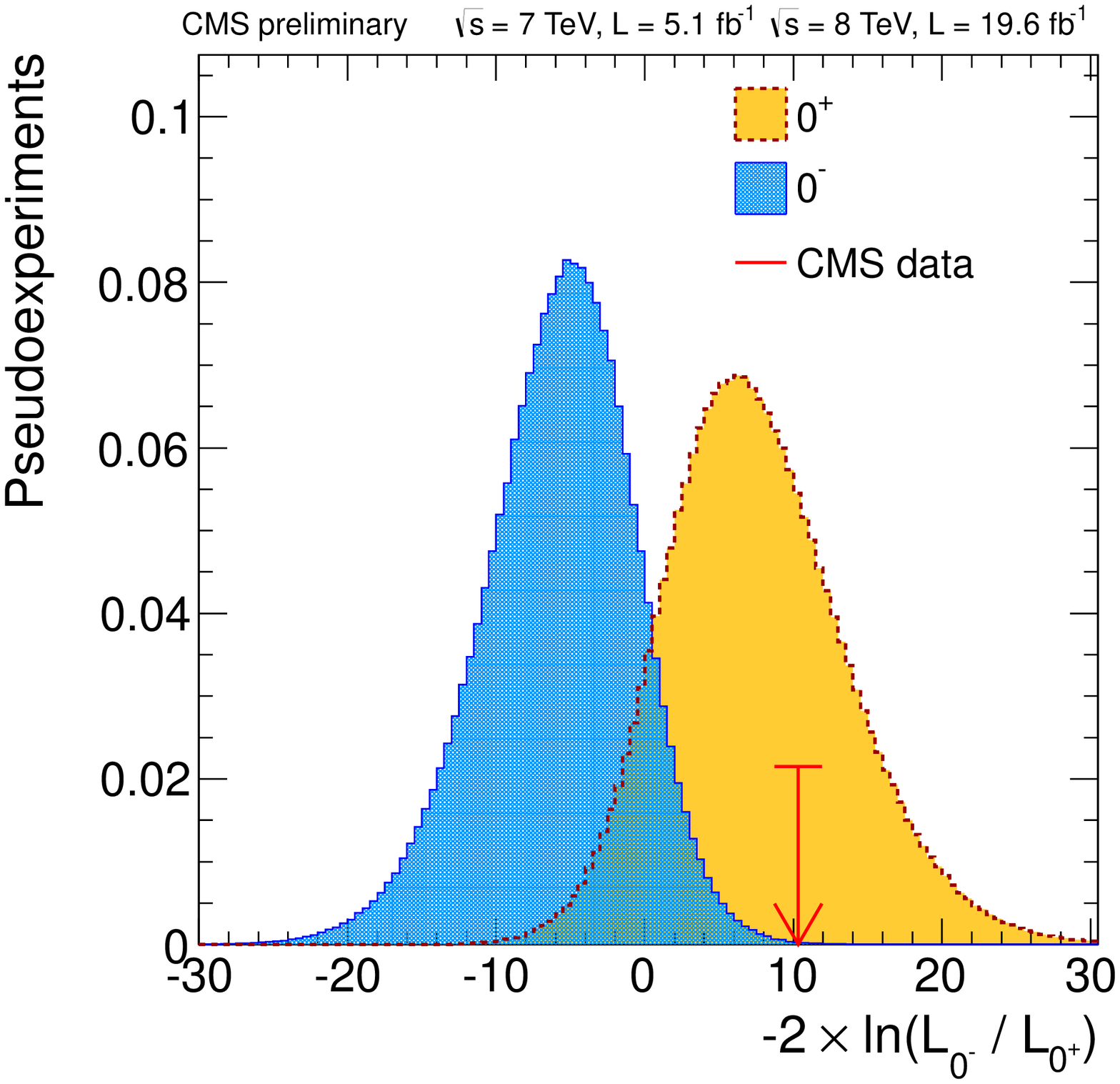}
\includegraphics[width=0.32\linewidth]{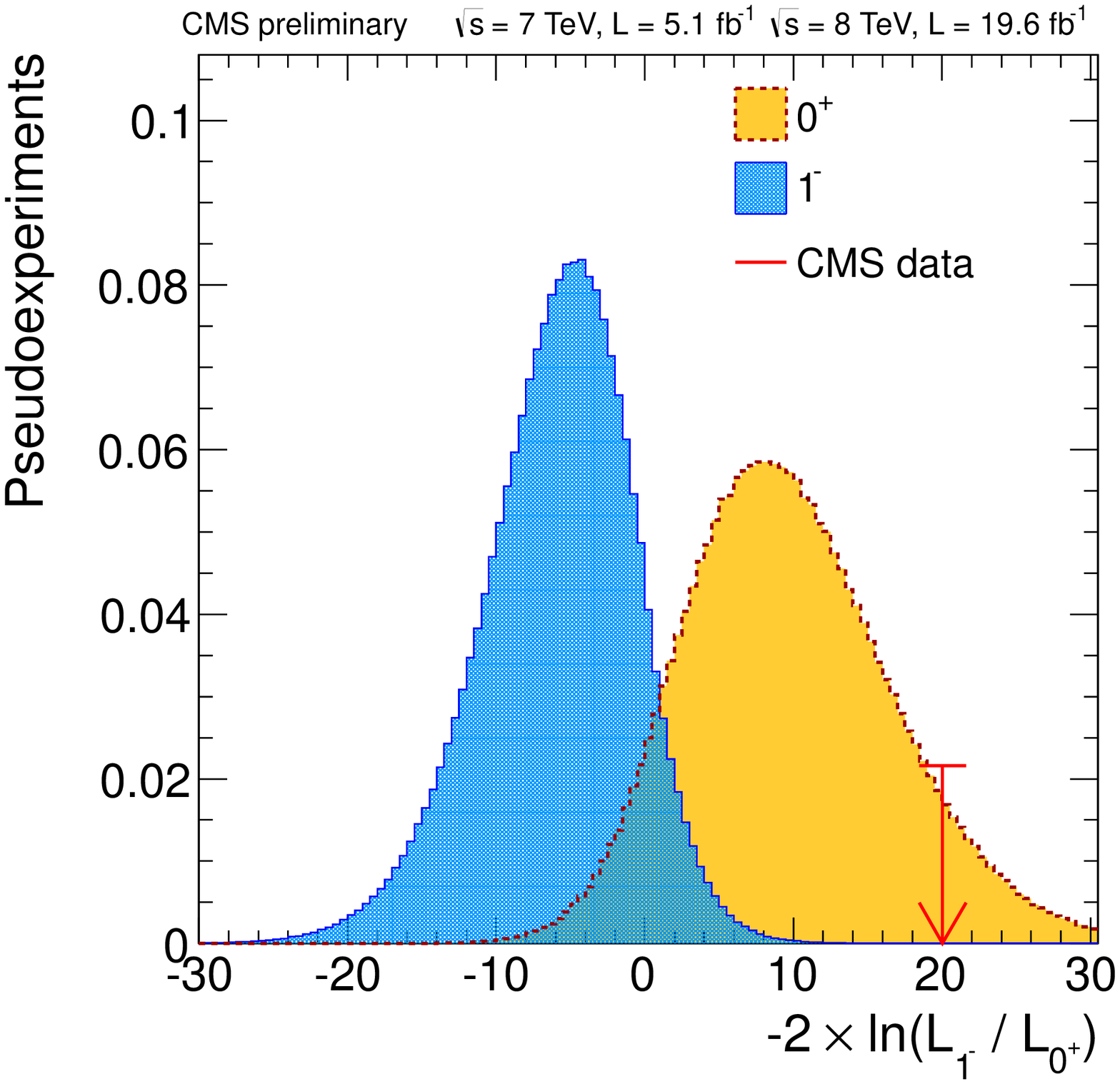}
\includegraphics[width=0.32\linewidth]{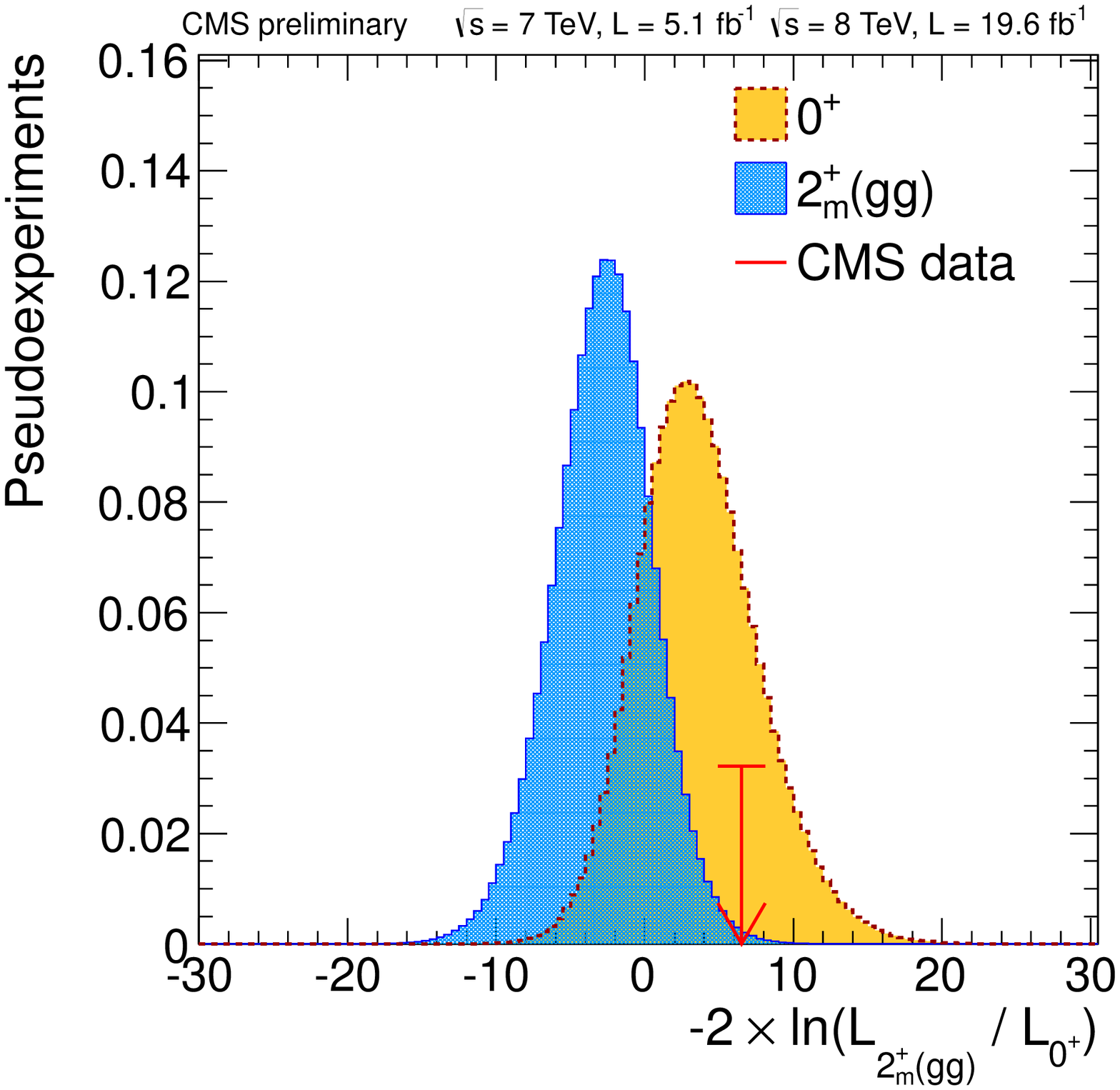}
}
\centerline{
\includegraphics[width=0.32\linewidth]{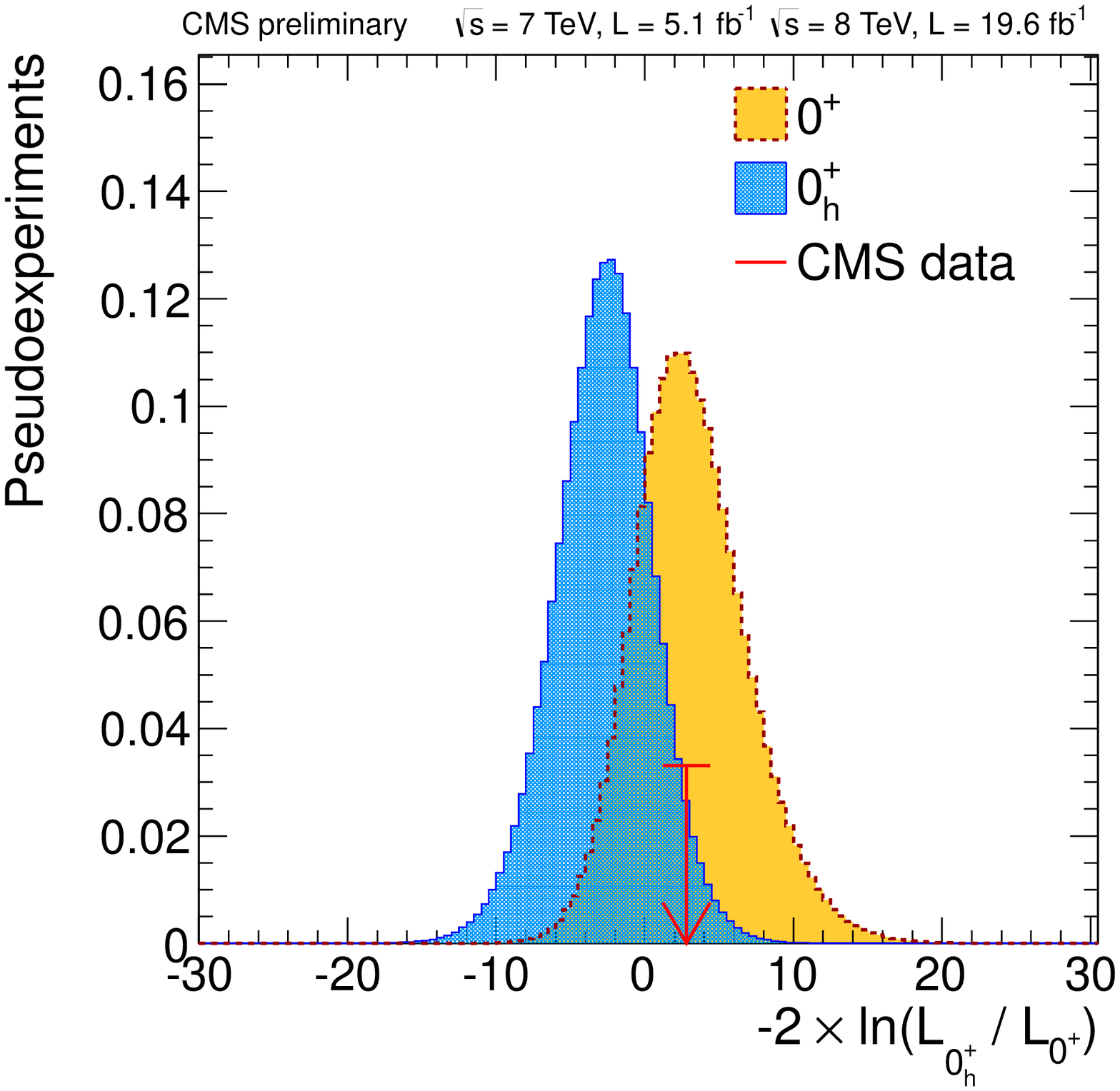}
\includegraphics[width=0.32\linewidth]{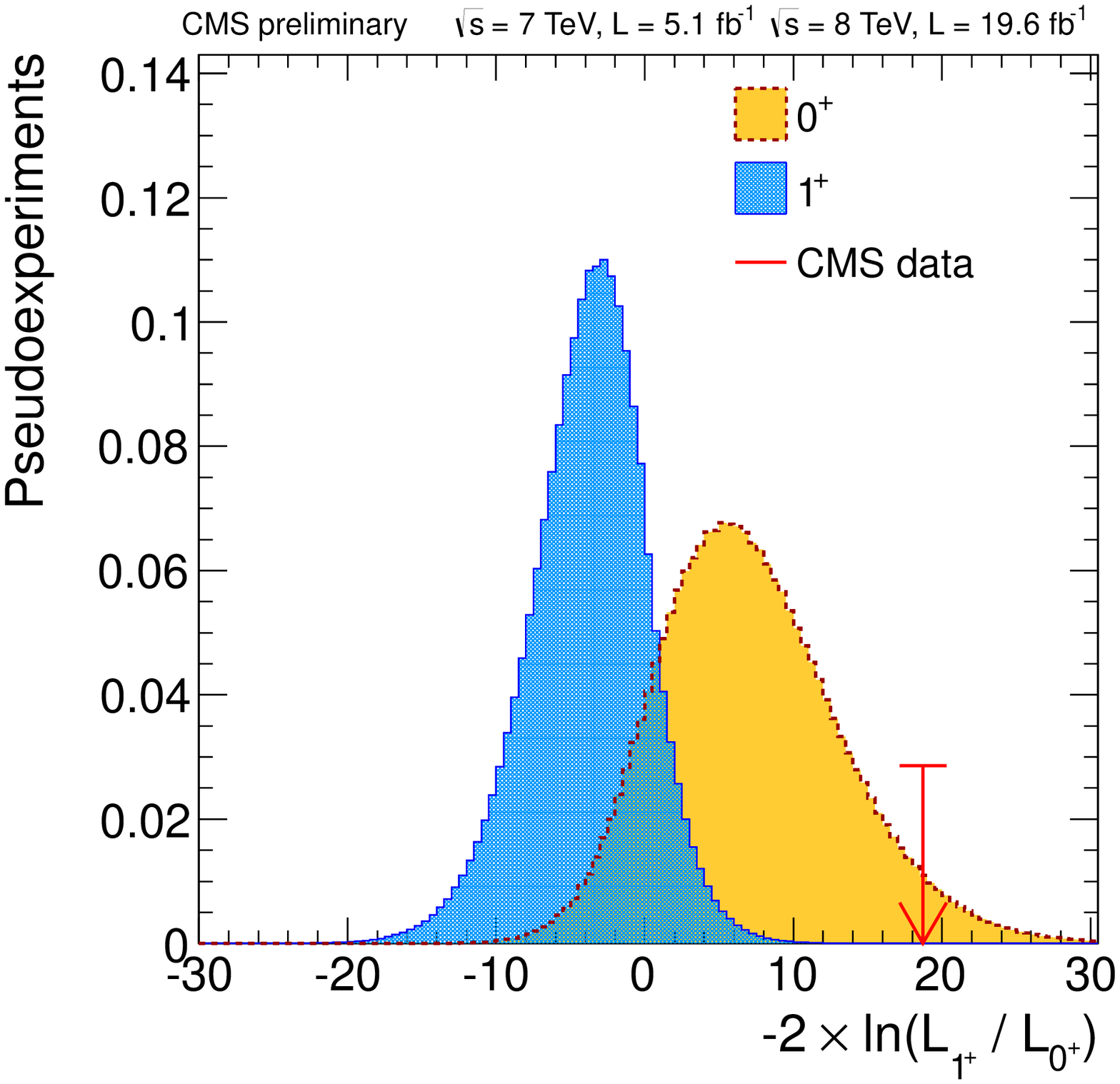}
\includegraphics[width=0.32\linewidth]{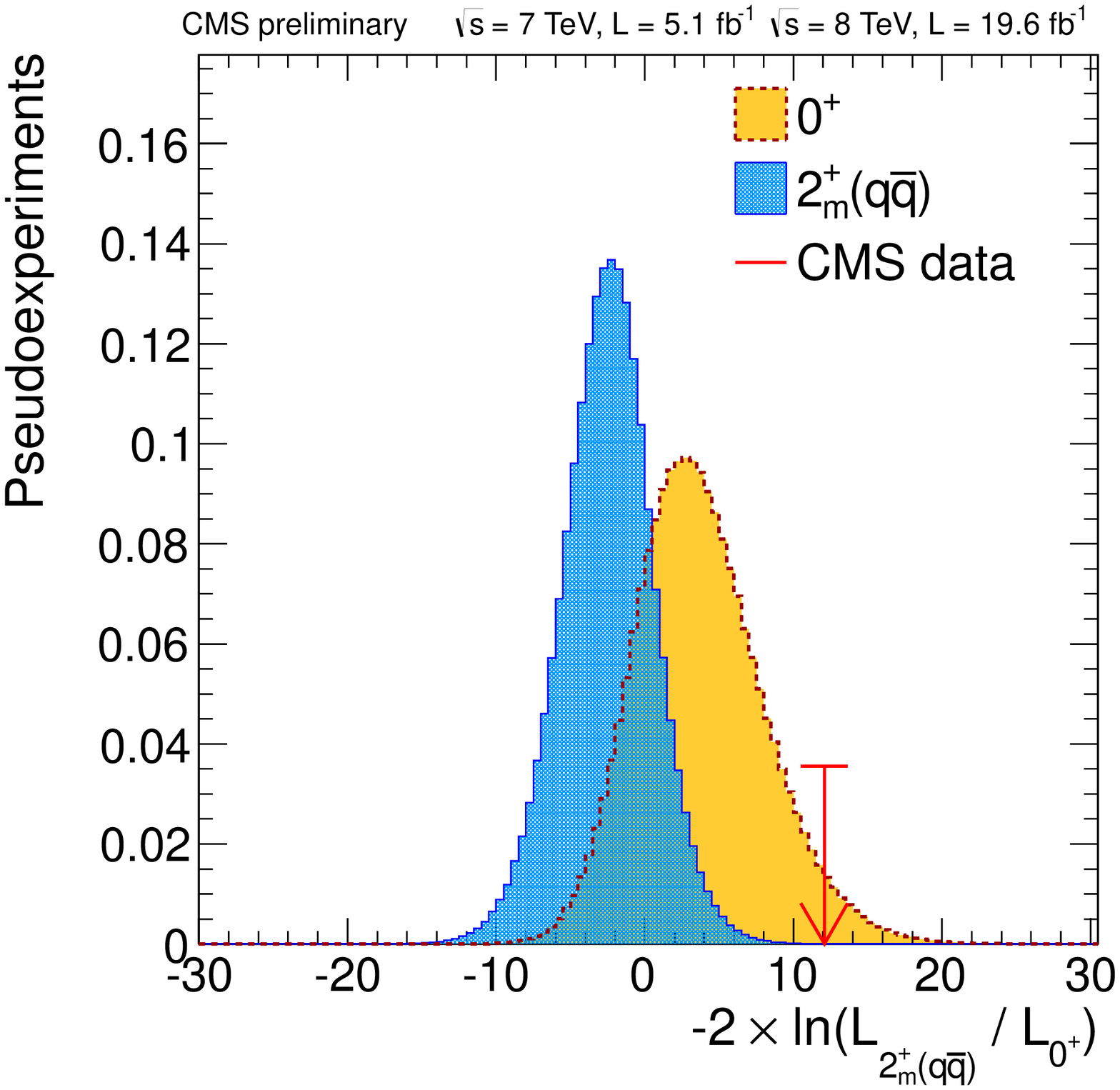}
}
\caption{
Post-fit model distribution of $q=-2{\ln({\cal L}_{J^P}/{\cal L}_{\rm SM})}$ for two signal types
($0^+$ histogram to the right and $J^P$ histogram to the left)
for $m_H =126$ GeV shown with a large number of generated experiments.
The arrow indicates the observed value.
Six alternative hypotheses are tested from top to bottom and left to right:
$J^P=0^-, 0_h^+,1^-, 1^+, 2_{~mgg}^+,  2_{~mq\bar{q}}^+$.
}
\label{fig:kd_jp_separation}
\end{center}
\end{figure}

The expected distribution of the test statistic under background plus signal
hypothesis is built by generating pseudoexperiments, assuming $m_{\PH} = 126 (125)$ GeV in the $\mathrm{ZZ} \to 4\ell$ ($\PW\PW \to \ell\nu\ell\nu$) channel. 
The expected distribution is computed for two scenarios:
i) the pre-fit model, where all
nuisance parameters are set to their default values before fitting the data and
the $\mu$ is set to $1$, and 
ii) the post-fit model, where all parameters are set to
their best-fit values when fitting the data.
We find the resutls from the two scenarios are consistent. 
Figure~\ref{fig:kd_jp_separation} shows the post-fit distributions as well as the observed values of the test statistic for the six hypothesis tests in the $\mathrm{ZZ} \to 4\ell$ channel. 

A $\CLs$ criterion is defined as the ratio of the probabilities to find, 
 under each of the hypotheses, values of the test statistic equal or larger than the
 one observed in the data: 
\begin{equation}
 CLs^\mathrm{obs.}  = P(\,q \geq q^\mathrm{obs.} \, | \,
 J^P \,)/ P(\,q \geq q^\mathrm{obs.} \,| \, SM \,).
\end{equation}
The expected separation is defined as the tail probability which is calculated at the value of $q$ where the tails of the two distributions have identical area.

The expected and observed results in  the $\mathrm{ZZ} \to 4\ell$ channel
 are summaried in Table~\ref{tab:jpmodels}. The data disfavor the alternative hypotheses $J^P$ with a CLs value in the range 0.1-10\%.

The observed result in the $\PH\to \PW\PW \to \ell\nu\ell\nu$ channel is that
 the data disfavor the  $2_\mathrm{m}^+$(gg) hypothesis with a CLs value 14\%, 
with the median expected 1.8 (2.4) $\sigma$ deviation from SM with post-fit (pre-fit) scenario.

\begin{table}[h]
\centering
\caption{
List of models used in analysis of spin-parity hypotheses corresponding to the pure states of the type noted in the $\mathrm{ZZ} \to 4\ell$ channel.
The expected separation is quoted for two scenarios corresponding to pre-fit model ($\mu$=1) and post-fit model.
The observed separation quotes consistency of the observation with the $0^+$ model or $J^P$ model,
and corresponds to the post-fit model. The last column quotes CL$_s$ criterion for the $J^P$ model.
}
\begin{tabular}{|l|c|c|c|c|c|c|}
\hline
 $J^P$ & production & comment & expect ($\mu$=1) &  obs. $0^+$  & obs. $J^P$ & CL$_s$  \\
\hline
$0^-$ & $gg\to X$ & pseudoscalar  &  2.6$\sigma$ (2.8$\sigma$)  & 0.5$\sigma$ & 3.3$\sigma$ &  0.16\%  \\
$0_h^+$ & $gg\to X$  & higher dim operators &  1.7$\sigma$ (1.8$\sigma$)  &  0.0$\sigma$ &  1.7$\sigma$ &  8.1\% \\
$2_{~mgg}^+$ & $gg\to X$  & minimal couplings &  1.8$\sigma$ (1.9$\sigma$)  &  0.8$\sigma$ &  2.7$\sigma$ &  1.5\% \\
$2_{~mq\bar{q}}^+$& $q\bar{q}\to X$   & minimal couplings &  1.7$\sigma$ (1.9$\sigma$)   &  1.8$\sigma$ &  4.0$\sigma$  &  $<$0.1\% \\
$1^-$& $q\bar{q}\to X$   & exotic vector &  2.8$\sigma$ (3.1$\sigma$) &  1.4$\sigma$ &  $>$$4.0$$\sigma$  &  $<$0.1\% \\
$1^+$& $q\bar{q}\to X$   & exotic pseudovector &  2.3$\sigma$ (2.6$\sigma$)  &  1.7$\sigma$  &  $>$$4.0$$\sigma$ &  $<$0.1\% \\
\hline
\end{tabular}
 \label{tab:jpmodels}
\end{table}

\section{Conclusions}
Combination results of the recently discovered boson are presented using data samples 
corresponding to integrated luminosities of up to 5.1 $\invfb$ at 7 TeV and 
up to 12.2 $\invfb$ at 8 TeV of proton-proton collisions collected with CMS experiment at LHC.   
The significance of the new boson is 6.9 $\sigma$ with mass measured to be 125.8 $\pm$ 0.4 (stat) $\pm$ 0.4 (syst). 
The event yields obtained by the different analyses targeting specific decay modes and production mechanisms are consistent with those 
predicted for the stand model (SM) Higgs boson. The best-fit signal strength for all channels combined,
 expressed in units of the SM Higgs boson cross section, is 0.88 $\pm$ 0.21 at the measured mass. 
The consistency of the couplins of the observed boson with those expected for the SM Higgs boson is tested in various ways, 
and no significant deviations are found. 
Results on the test of different spin-parity hypotheses of the observed boson are also shown, 
but with updated data samples corresponding to integrated luminosities of 5.1 $\invfb$ at 7 TeV and 
19.6 $\invfb$ at 8 TeV in two channels H $\rightarrow$ WW $\rightarrow~2\ell2\nu$ and H $\rightarrow$ ZZ $\rightarrow~4\ell$ separately.  
Under the assumption that the observed boson has spin 0 and positive parity, 
the pure scalar hypothesis is found to be consistent with the observed boson when compared to other tested spin-parity hypotheses.
 The data in the H $\rightarrow$ ZZ $\rightarrow~4\ell$ channel disfavor the pseudo-scalar hypothesis $0^-$ with a CLs value of 0.16$\%$, disfavor the pure spin-2 hypothesis of a narrow
resonance with the minimal couplings to the vector bosons with a CLs value of 1.5$\%$, and disfavor the pure spin-1 hypothesis with even smaller CLs value.

\section*{References}
\bibliographystyle{unsrt} 
\bibliography{moriond_final}

\end{document}